\RequirePackage{lineno}
\documentclass[preprint,amsmath,amssymb,superscriptaddress,unsortedaddress]{revtex4}

\usepackage{color}
\usepackage{newfloat}
\usepackage{amsmath}
\usepackage{graphicx}
\usepackage{setspace}
\usepackage{ulem}

\begin{document}

\title{Devil's staircase transition of the electronic structures in CeSb}

\author{Kenta~Kuroda\textsuperscript{*}}
\affiliation{ISSP, University of Tokyo, Kashiwa, Chiba 277-8581, Japan}

\author{Y.~Arai} 
\affiliation{ISSP, University of Tokyo, Kashiwa, Chiba 277-8581, Japan}

\author{N.~Rezaei} 
\affiliation{Department of Physics, Isfahan University of Technology, 84156-83111 Isfahan, Iran}

\author{S.~Kunisada} 
\affiliation{ISSP, University of Tokyo, Kashiwa, Chiba 277-8581, Japan}

\author{S.~Sakuragi} 
\affiliation{ISSP, University of Tokyo, Kashiwa, Chiba 277-8581, Japan}

\author{M.~Alaei} 
\affiliation{Department of Physics, Isfahan University of Technology, 84156-83111 Isfahan, Iran}

\author{Y.~Kinoshita} 
\affiliation{ISSP, University of Tokyo, Kashiwa, Chiba 277-8581, Japan}

\author{C.~Bareille} 
\affiliation{ISSP, University of Tokyo, Kashiwa, Chiba 277-8581, Japan}

\author{R.~Noguchi} 
\affiliation{ISSP, University of Tokyo, Kashiwa, Chiba 277-8581, Japan}

\author{M.~Nakayama} 
\affiliation{ISSP, University of Tokyo, Kashiwa, Chiba 277-8581, Japan}

\author{S.~Akebi} 
\affiliation{ISSP, University of Tokyo, Kashiwa, Chiba 277-8581, Japan}

\author{M.~Sakano} 
\affiliation{ISSP, University of Tokyo, Kashiwa, Chiba 277-8581, Japan}
\affiliation{Department of Applied Physics, University of Tokyo, 7-3-1 Hongo Bunkyo-ku, Tokyo 113-8656, Japan}

\author{K.~Kawaguchi} 
\affiliation{ISSP, University of Tokyo, Kashiwa, Chiba 277-8581, Japan}

\author{M.~Arita} 
\affiliation{Hiroshima Synchrotron Center, Hiroshima University, Higashi-Hiroshima 739-0046, Japan}

\author{S.~Ideta} 
\affiliation{UVSOR Facility, Institute for Molecular Science, Okazaki 444-8585, Japan}

\author{K.~Tanaka} 
\affiliation{UVSOR Facility, Institute for Molecular Science, Okazaki 444-8585, Japan}

\author{H.~Kitazawa} 
\affiliation{National Institute for Materials Science, 1-2-1 Sengen, Tsukuba 305-0047, Japan}

\author{K.~Okazaki} 
\affiliation{ISSP, University of Tokyo, Kashiwa, Chiba 277-8581, Japan}

\author{M.~Tokunaga} 
\affiliation{ISSP, University of Tokyo, Kashiwa, Chiba 277-8581, Japan}

\author{Y.~Haga} 
\affiliation{Advanced Science Research Center, Japan Atomic Energy Agency, Tokai, Ibaraki 319-1195, Japan}

\author{S.~Shin} 
\affiliation{ISSP, University of Tokyo, Kashiwa, Chiba 277-8581, Japan}

\author{H.~S.~Suzuki} 
\affiliation{ISSP, University of Tokyo, Kashiwa, Chiba 277-8581, Japan}

\author{R.~Arita} 
\affiliation{Department of Applied Physics, University of Tokyo, 7-3-1 Hongo Bunkyo-ku, Tokyo 113-8656, Japan}
\affiliation{RIKEN Center for Emergent Matter Science (CEMS), 2-1 Hirosawa, Wako, Saitama 351-0198, Japan}

\author{Takeshi~Kondo} 
\affiliation{ISSP, University of Tokyo, Kashiwa, Chiba 277-8581, Japan}

\date{\today}

\maketitle

\noindent\\
\section*{Abstract}
\textbf{
Solids with competing interactions often undergo complex phase transitions with a variety of long-periodic modulations.
Among such transition, devil's staircase is the most complex phenomenon, and for it, CeSb is the most famous material, where a number of the distinct phases with long-periodic magnetostructures sequentially appear below the N{\'{e}}el temperature.
An evolution of the low-energy electronic structure going through the devil's staircase is of special interest, which has, however, been elusive so far despite the 40-years of intense researches.
Here we use bulk-sensitive angle-resolved photoemission spectroscopy and reveal the devil's staircase transition of the electronic structures.
The magnetic reconstruction dramatically alters the band dispersions at each transition.
We moreover find that the well-defined band picture largely collapses around the Fermi energy under the long-periodic modulation of the transitional phase, while it recovers at the transition into the lowest-temperature ground state.
Our data provide the first direct evidence for a significant reorganization of the electronic structures and spectral functions occurring during the devil's staircase.
}

\newpage
%
%
\section*{Introduction}
An emergence of spatially modulated structures such as ordered magnetostructures~\cite{Chattopadhyay_science1994}, charge-density waves~\cite{Gruner_rmp1988} and orbital orders~\cite{Tokura_science2000} is very common in condensed matters.
Their modes can be characterized by a particular periodicity and symmetry, both of which are often related to an underlying electronic instability~\cite{Monceau_advance2012} since the quasiparticle excitation should be subjected to the boundary newly generated, in addition to that of the periodic crystal lattice.
When several different modes compete with each other, the system is influenced by the frustration and consequently exhibits rather complex properties~\cite{Bak_rpp1982}: in particular, a long-period modulation tends to be build up causing exceedingly complex phase transitions.
The most complex phenomenon is known as ``devil's staircase''~\cite{Bak_PhysToday1986}, which was first discovered in the magnetically ordered CeSb more than 40 years ago~\cite{Bak_prl1979}.

At zero-field, the seven phases with different magnetostructures sequentially appear one after another with decreasing temperature below N{\'{e}}el temperature ($T_{\rm{N}}$) $\sim$17~K~\cite{Mignod_prb1977,Fischer_jpc1978,Mignod_jpc1980,Mignod_jmmm1983} (Fig.~\ref{fig1}\textbf{a}).
They comprise stacking of the Ising-like ferromagnetic (001) planes with the moment of $\sim$2$\mu_{\rm{B}}$ perpendicular to the plane, and differ in the sequence of the square wave modulation ($q$).
The ground state is established below $\sim$8~K ($T_{\rm{AF}}$) as the antiferromagnetic (AF) phase with the double-layer stacking of the ferromagnetic planes.
As a symbolic property of the devil's staircase, various transitional phases show up just below the $T_{\rm{N}}$, where the double-layer AF modulation is periodically locked by the rest paramagnetic (P) layer.
Hereafter we call these transitional phases antiferro-paramagnetic (AFP) phases.
The transition from the AFP5 to AFP6 phase occurs without a clear change of the $q$ vector~\cite{Mignod_prb1977} but with a large variation of entropy~\cite{Mignod_jpc1980}.

The very existence of the devil's staircase poses questions of how the modulation of the Ce 4$f$ states manifests itself in electronic structures; this has been a longtime issue to be addressed to understand the mechanism of the fascinating phenomenon.
So far, the paramagnetic electronic structure with the semimetalic feature was well documented in detail~\cite{Hasegawa_jpsj85,Kasuya_jmmm74,Kuroda_prl2018,Oinuma_prb2018}: two hole pockets of Sb 5$p$ at $\rm{\Gamma}$ point and an electron pocket of Ce 5$d$ at X point (Fig.~\ref{fig1}\textbf{b}).
Under the AF modulation, these bands should be transformed from a cubic to a tetragonal structure in the reduced Brillouin zone (BZ) (Fig.~\ref{fig1}\textbf{c}), and the consequent band-folding is theoretically expected to generate hybridization gaps at the Fermi energy ($E_{\rm{F}}$)~\cite{Ishiyama_jpsj03}.
In addition, theory has also pointed out importance of interactions between the localized 4$f$ states and mobile electrons~\cite{Siemann_prl1980,Takahashi_jpc1985,Kioussis_prb1988,Kasuya_physics1993}, particularly intralayer hybridizations with the Sb 5$p$ electrons ($p$-$f$ mixing)~\cite{Takahashi_jpc1985, Kasuya_physics1993}, to explain the anomalous properties of CeSb below $T_{\rm{N}}$.
While the 4$f\Gamma_{7}$ state is the ground state of the Ce$^{3+}$ ions with 4$f^1$ state in P phase, through the quadrupolar $p$-$f$ mixing, the cruciform 4$f\Gamma_8$ orbit close to a fully polarized state of $J_{z}$ = $|{\pm}$5/2$>$ is favored as a ground state in the ferromagnetic plane~\cite{Takahashi_jpc1985,Kasuya_physics1993}. 
This cruciform 4$f\Gamma_8$ orbit is responsible for the strong magnetic anisotropy~\cite{Mignod_jmmm1985}, cubic-to-tetragonal lattice distortions along 4$f$ moment~\cite{hulliger_jltp1975,Iwasa_jpsj1999,Iwasa_prl2002} and a dramatic change in the electron transport~\cite{Mori_jap1991, SettaiJPSJ1994,Ye_prb2018,Xu_nacom2019}, all of which coincide with the devilish behavior. 
Momentum-resolved spectral function obtained by angle-resolved photoemission spectroscopy (ARPES)~\cite{Damascelli_rmp2003} should be an ideal probe to reveal the compatible transitions of the electronic structure and spectral transfer related to underlying electronic instability~\cite{Kampf_jpcs1995}.
However, in earlier ARPES reports~\cite{Kumigashira_prb1997,Ito_physica2004,Jangeaat_scienceAd2019,Takayama_jpsj2009},
no systematic investigation over the devil's staircase particularly for transitional AFP phases has been reported, and hence no correspondence for the devil's staircase has been achieved so far.

In this study, we use laser-based ARPES to directly visualize the electronic structures of CeSb developing through the devil's staircase. 
In contrast to the previous reports~\cite{Kumigashira_prb1997,Ito_physica2004,Jangeaat_scienceAd2019,Takayama_jpsj2009}, the high-energy resolution and bulk-sensitivity achieved by utilizing a low-energy laser source ($h{\nu}$=7~eV)~\cite{ShimojimaJPSJ2015} enables us to obtain high-quality spectra, which now unravels the significant reconstruction of the itinerant bands.
The response of conducting bands to the 4$f$ order is sensitively changed at each distinct transition of the devil's staircase, and it exposes the strong electronic anisotropy across $T_{\rm{N}}$.
Interestingly, the well-defined band picture with coherent quasiparticles is verified in AF phase in the ground state ($T<T_{\rm{AF}}$), whereas it partially collapses in momentum space around $E_{\rm{F}}$ under the emergence of the long-periodic modulation in AFP6 phase.\\

\section*{Results}
\noindent
{\textbf{Spatial mapping for tetragonal domains.}}\\
Let us start by showing that the tetragonal transition across $T_{\rm{N}}$ forms multiple domains randomly distributed in the crystal at zero-field (the definitions of the domain are described in Supplementary-Fig.~1).
It is necessary to discriminate these tetragonal domains in measurements by macroscopic probes such as ARPES; otherwise, the mixture of the signatures from the domains easily obscures the intrinsic electronic properties.
While the presence of such domains was previously argued~\cite{Mignod_prb1977,Fischer_jpc1978}, no direct observation for it has been so far reported.
We, therefore, use the polarizing microscope~\cite{Ishikawa_prb1999} and spatially map the tetragonal domains on the cleaved (001) surface in Fig.~\ref{fig2}\textbf{e}.
The optical birefringence is clearly observed in the difference between the microscope images obtained above and below $T_{\rm{N}}$ (Figs.~\ref{fig2}\textbf{c} and \ref{fig2}\textbf{d}, respectively).
In our optical geometry, the red- and blue-colored areas should exhibit the distribution of the domain with the 4$f$ moment lying along either [100] or [010] (we call hereafter $ab$-domain, Fig.~\ref{fig2}\textbf{a}), while the white-colored area indicating no birefringence effect corresponds to the regions for the other domain with the moment along [001] ($c$-domain, Fig.~\ref{fig2}\textbf{b}).
Since the optical property is tied to the electronic structure, the observed birefringence already indicates a presence of the anisotropic reconstruction below $T_{\rm{N}}$.\\

\noindent
{\textbf{Impacts of the 4$f$ order on the electronic structure.}}\\
In order to directly observe the reconstruction of the electronic bands due to the 4$f$ order, we employed bulk-sensitive laser-ARPES with a spot size of less than 50 $\rm{\mu}$m, sufficiently small to selectively observe the different tetragonal domains (circle in Fig.~\ref{fig2}\textbf{e}).
Our laser with a low $h\nu$ sensitively detects the bulk band dispersions at $k_{z}{\sim}$0.2~${\rm{\AA}}^{-1}$ (Fig.~\ref{fig3}\textbf{a}, Supplementary Fig.~2).
Accordingly, the laser-ARPES map of P phase displays two hole-like parabolas near $E_{\rm{F}}$ (Fig.~\ref{fig3}\textbf{g}), showing a good agreement with the DFT bands of Sb 5$p$ hole-bands (Fig.~\ref{fig3}\textbf{j}).

\begin{figure}
\centering
\includegraphics[width=\textwidth]{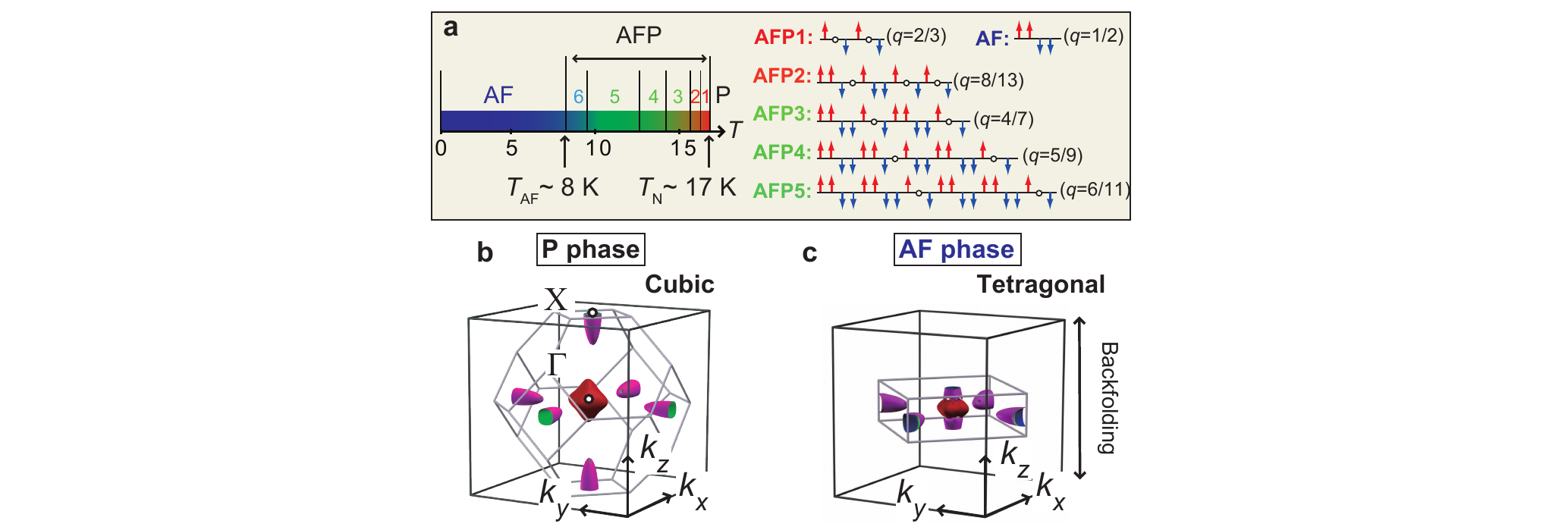}
\caption{\label{fig1}
\textbf{Magnetostructures and the reconstruction of Fermi surfaces.}
(\textbf{a}) Various magnetostructures at zero-field below $T_{\rm{N}}\sim$17~K, which differ in the stacking sequence of the Ising-like ferromagnetic (001) planes (red and blue arrows) with the square wave modulation ($q$)~\cite{Mignod_prb1977,Fischer_jpc1978}.
Antiferromagnetic (AF) phase with the double-layer modulation is most favored at low-temperature below $T_{\rm{AF}}\sim$8~K while various antiferroparamagnetic (AFP) phases appear as transitional phases.
The cruciform 4$f{\Gamma}_{8}$ state is the ground state of the 4$f$ level in the ferromagnetic planes while the 4$f{\Gamma}_{7}$ ground state remains in the paramagnetic (P) planes (circles)~\cite{Iwasa_jpsj1999,Iwasa_prl2002}. 
The crystal lattice is slightly distorted below $T_{\rm{N}}$ with a shrink along the magnetic moment and the $q$ direction~\cite{hulliger_jltp1975}.
The detailed magnetostructure of AFP6 phase has not been determined yet~\cite{Mignod_jpc1980}.
(\textbf{b}) Fermi surfaces (FSs) in Brillouin zone (BZ) of the fcc lattice and (\textbf{c}) the folded FSs in the reduced tetragonal BZ under the AF modulation ($q$=1/2).
}
\end{figure}

\begin{figure*}
\centering
\includegraphics[width=\textwidth]{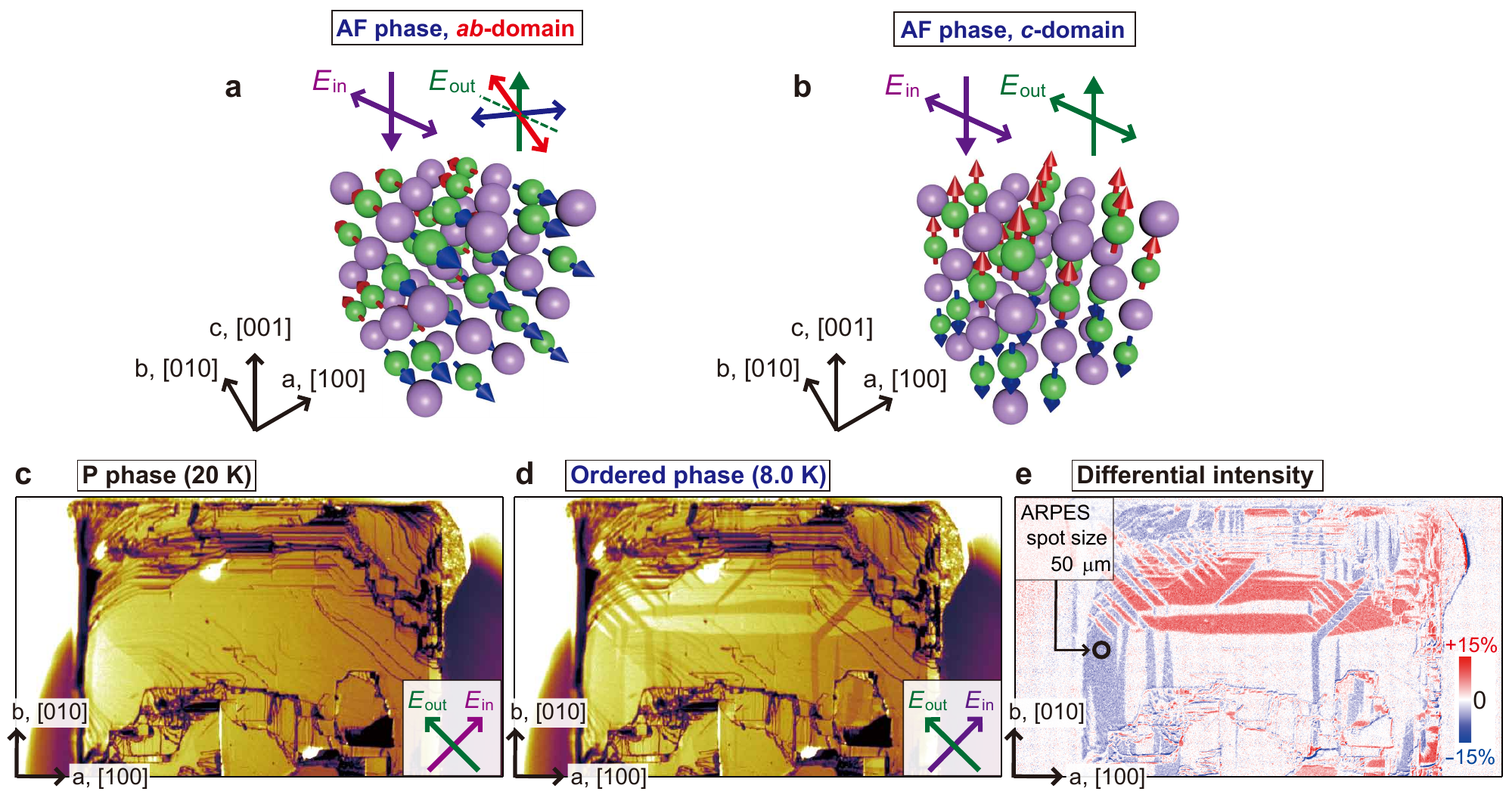}
\caption{\label{fig2}
\textbf{Polarizing microscope images of CeSb.}
The uniaxial magnetostructures with different orientation of the 4$f$ moment along [100] or [010] (\textbf{a}, $ab$-domain) and along [001] (\textbf{b}, $c$-domain).
The double arrows indicate the optical principal axis of the incident light (purple arrow) and reflected light (blue, red and green arrows). 
Polarizing microscope images of the cleaved (001) surface at (\textbf{c}) P phase at 20~K and (\textbf{d}) the ordered phase at 8~K.
These images were taken under crossed Nicols configuration that is sensitive to $ab$-domain (red and blue areas) but insensitive to $c$-domain.
The differential image presented in (\textbf{e}) to display the domain distributions.
The size of the domain is sufficiently large for the spot size of the laser light in our laser-ARPES experiments (circle). 
}
\end{figure*}

\begin{figure*}
\centering
\includegraphics[width=\textwidth]{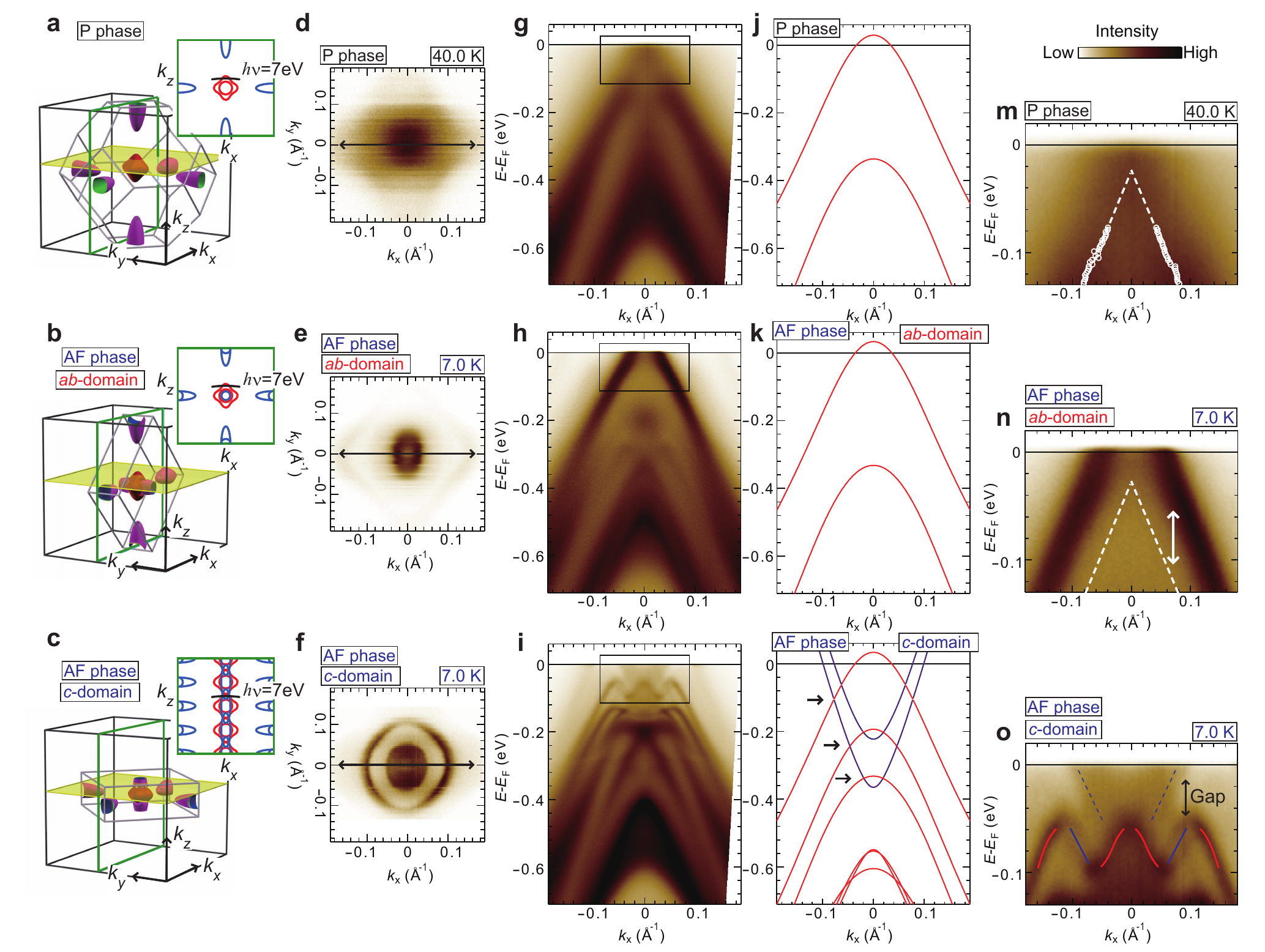}
\caption{\label{fig3}
\textbf{Dramatic electronic reconstruction of ordered CeSb in AF phase.}
(\textbf{a}) Three-dimensional FSs for P phase, and the folded one for (\textbf{b}) $ab$- and (\textbf{c}) $c$-domain according to the AF modulation ($q$=1/2).
The yellow colored planes correspond to the $k_x$-$k_y$ sheets at $k_z$=0.2~$\text{\AA}^{-1}$ detected by our laser-ARPES with $h\nu$=7~eV (Supplementary Fig.~2).
The insets schematically illustrate the (red line) hole and (blue line) electron pockets in $k_z$-$k_x$ plane (green frames in \textbf{a-c}) with (black line) the $k_x$ cut line at $k_z$=0.2~$\text{\AA}^{-1}$ for our laser-ARPES.    
The FS maps experimentally observed at (\textbf{d}) 40~K (P phase) and 7.0~K (AF phase) for different domains, (\textbf{e}) $ab$- and (\textbf{f}) $c$-domain, which were taken at the different positions of the same cleaved surface.
In \textbf{d}, we also observe the blurred photoelectron intensity outside the main signal due to the $k_z$ broadening effect~\cite{Kuroda_prl2018, Strocov03jesrp}.
(\textbf{g-i}) The $E$-$k_x$ maps cut along the $k_x$ line denoted by arrows in \textbf{d-f}.
(\textbf{j-l}) The DFT bands cut along the comparable $k_x$ line with \textbf{g-i}.
The DFT bands in AF phase was computed with the backfolding of the bands of P phase into the deduced BZ (see \textbf{b, c}).
The overall bands are shifted in energy with $-0.1$~eV~\cite{Kuroda_prl2018} to compare the experimental results. 
The dispersions of the hole and electron bands are indicated by red and blue colored lines, respectively.
The arrows in \textbf{l} highlight the band crossings which allow the hybridization.  
(\textbf{m-o}) the magnified laser-ARPES images within the $E$-$k_x$ windows indicated by rectangles in \textbf{g-i}.
In \textbf{m, n}, the band dispersion for P phase (dashed lines) was deduced by the peak position of the momentum distribution curves (circles in \textbf{m}).
In \textbf{o}, the dispersions of the hole and electron bands are guided by red and blue lines, respectively.
}
\end{figure*}

We now show that the itinerant bands are significantly reconstructed by the emergence of the 4$f$ order and transformed to those with anisotropic symmetry due to the backfolding of the BZ (see Fig.~\ref{fig1}\textbf{c}).
Consequently, our laser-ARPES can determine, by resolving $ab$- and $c$-domains, the folded FSs on the different momentum sheets in their reduced BZ (yellow planes in Figs.~\ref{fig3}\textbf{b} and \ref{fig3}\textbf{c}).
The clear signatures for the reconstruction are presented in the laser-ARPES maps (Figs.~\ref{fig3}\textbf{h} and \ref{fig3}\textbf{i}) obtained at 7.0~K below $T_{\rm{AF}}$ (AF phase) at the different positions of the same cleaved surface.
Even just a glance at Fig.~\ref{fig3}\textbf{i} reveals a huge transformation in the electronic structure due to the emergence of a number of the folded bands.
In sharp contrast, the band transformation is very weak in Fig.~\ref{fig3}\textbf{h}.
This contrastive feature is qualitatively consistent with the DFT calculations separately considering $ab$- and $c$-domains (Figs.~\ref{fig3}\textbf{k} and \ref{fig3}\textbf{l}, respectively) with the backfolding of bands into each reduced BZ.
In particular, the impact of the backfolding is more effective at the measured $k_z$ plane for $c$-domain (inset of Fig.~\ref{fig3}\textbf{c}) than that for $ab$-domain (inset of Fig.~\ref{fig3}\textbf{b}).
Correspondingly, the folded electron bands are observed only for $c$-domains (Fig.~\ref{fig3}\textbf{i}, and blue lines in Fig.~\ref{fig3}\textbf{l}).
These results thus unveil the dramatic reconstruction of the itinerant bands, which fully converts the electronic periodicity from cubic to tetragonal symmetry.
In passing, we note, while the 3$^{\rm{rd}}$ domain should exist as a counterpart of $ab$-domain (see Fig.~\ref{fig2}\textbf{a}, also in Supplementary Fig.~1), it cannot be disentangled by our laser-ARPES because the discrepancy of the dispersions along $k_x$ and $k_y$ lines is rather weak at the measured $k_z$ plane (Fig.~\ref{fig3}\textbf{e}, Supplementary Fig.~3).

In Figs.~\ref{fig3}\textbf{m-o}, we show the enlarged laser-ARPES images of the rectangular regions in Figs.~\ref{fig3}\textbf{g-i}, to further examine the electronic reconstruction in the vicinity of $E_{\rm{F}}$.
Our data reveal that the backfolding of the Ce 5$d$ electron-bands causes hybridizations with the Sb 5$p$ hole-bands ($p$-$d$ mixing) where they cross each other (arrows in Fig.~\ref{fig3}\textbf{l}), leading to a significant modification in their band structures.
As a consequence of the $p$-$d$ mixing, the $zigzag$-shaped dispersion is observed at $E-E_{\rm{F}}\sim-$0.08~eV in our data of $c$-domain (Fig.~\ref{fig3}\textbf{o}). 
This band dispersion is characterized to be relatively flat, opening a hybridization gap larger than 50~meV below $E_{\rm{F}}$ (arrow in Fig.~\ref{fig3}\textbf{o}).
Such reconstruction should reduce the electronic energy, which is favorable in stabilizing the double-layer modulation of AF phase~\cite{Ishiyama_jpsj03}.

The early ARPES study confirmed an increment of the hole pocket volume in the ordered phase~\cite{Kumigashira_prb1997}, which was theoretically attributed to the enhanced quadrupole interaction between the Sb 5$p$ state and the cruciform 4$f\Gamma_8$ orbit~\cite{Takahashi_jpc1985,Kasuya_physics1993}.
Our data of $ab$-domain well captures this effect: the main hole-band in AF phase shifts upward in energy with respect to that in P phase (white lines in Figs.~\ref{fig3}\textbf{m} and \ref{fig3}\textbf{n}).
Consequently, we clearly observe a Fermi pocket at the measured $k_z$ plane in AF phase (Fig.~\ref{fig3}\textbf{e}).

A more astonishing finding in our data is that the spectral weight of the folded bands is very large and present in a wide energy range (Fig.~\ref{fig3}\textbf{i}).
The strong fingerprint for the folded bands, captured by our bulk-sensitive laser-ARPES, largely differs from the results obtained by previous surface-sensitive ARPES measurements~\cite{Kumigashira_prb1997,Ito_physica2004,Jangeaat_scienceAd2019}.
Moreover, our result contrasts with usual framework considered with long-range order models~\cite{Kampf_prb1990}.
In these, the spectral transfer relies on the coupling strength with the modulation field, which is generally too weak to complete the transfer of whole spectral weight~\cite{Voit_science2000}. 
Apparently, the dramatic behavior seen in CeSb indicates that the itinerant carriers sufficiently feel the periodicity and symmetry of the ordered 4$f$ states.
The strong coupling deduced from the spectral property in our data cannot be explained by the magnetic boundary which often only generates a weak signature of shadow bands in ARPES~\cite{Schafer_prl1999,Li_naturecom2013,Sobota_prb2013,Wallauer_prb2015}.
The observed band-folding with intensive spectral weight is likely induced by the underlying arrangement of the cruciform 4$f\Gamma_8$ orbit~\cite{Kasuya_physics1993,Mignod_jmmm1985,Iwasa_jpsj1999,Iwasa_prl2002} in ordered CeSb, which can electronically couple to the itinerant carriers and thus directly influence the electron motions.
This experimental hallmark elucidates the dramatic change of the electronic transport during the devil's staircase~\cite{Mori_jap1991,SettaiJPSJ1994,Ye_prb2018,Xu_nacom2019}.

\begin{figure*}
\centering
\includegraphics[width=\textwidth]{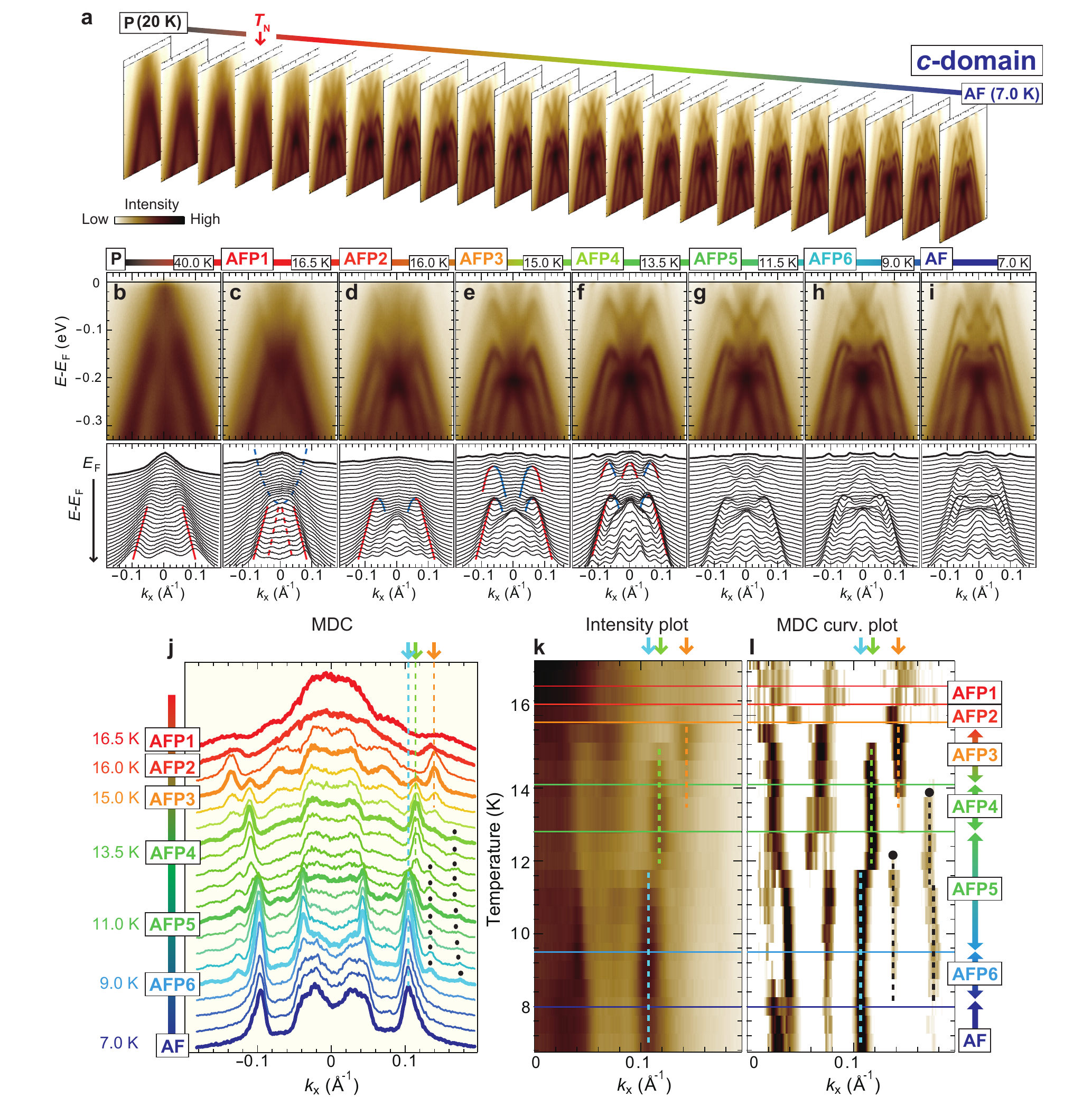}
\caption{\label{fig4}
\textbf{Devil's staircase evolution of the electronic structures.}
(\textbf{a}) Temperature evolution of the laser-ARPES images for $c$-domain from 20~K (P phase) to 7.0~K (AF phase) by a 0.5~K step (see also Supplementary movie~1).
(\textbf{b-i}) Laser-ARPES images around $E_{\rm{F}}$ at each phase through the devil's staircase, and the corresponding momentum distribution curves (MDCs).
The MDCs at $E_{\rm{F}}$ are highlighted with bold lines. 
Representative changes of the band dispersions by temperature are guided by colored lines. 
(\textbf{j}) Temperature evolution of the MDCs at $E_{\rm{F}}$ taken from the ARPES images shown in \textbf{a}.
The MDCs at each phase in \textbf{b-i} are highlighted with bold lines.
(\textbf{k, l}) Its two-dimensional map and the corresponding curvature plot~\cite{Zhang_rsi11} for $k_{x}>$0 to clearly visualize the appearances/disappearances of the quasiparticle peaks, corresponding to the devil's staircase transitions. 
The colored solid lines are the transition temperature previously determined by specific heat measurement~\cite{Mignod_jpc1980}. 
The colored arrows, circles and dashed lines indicate the representative peaks in the different phases.
}
\end{figure*}

\begin{figure*}
		\centering
\includegraphics[width=\textwidth]{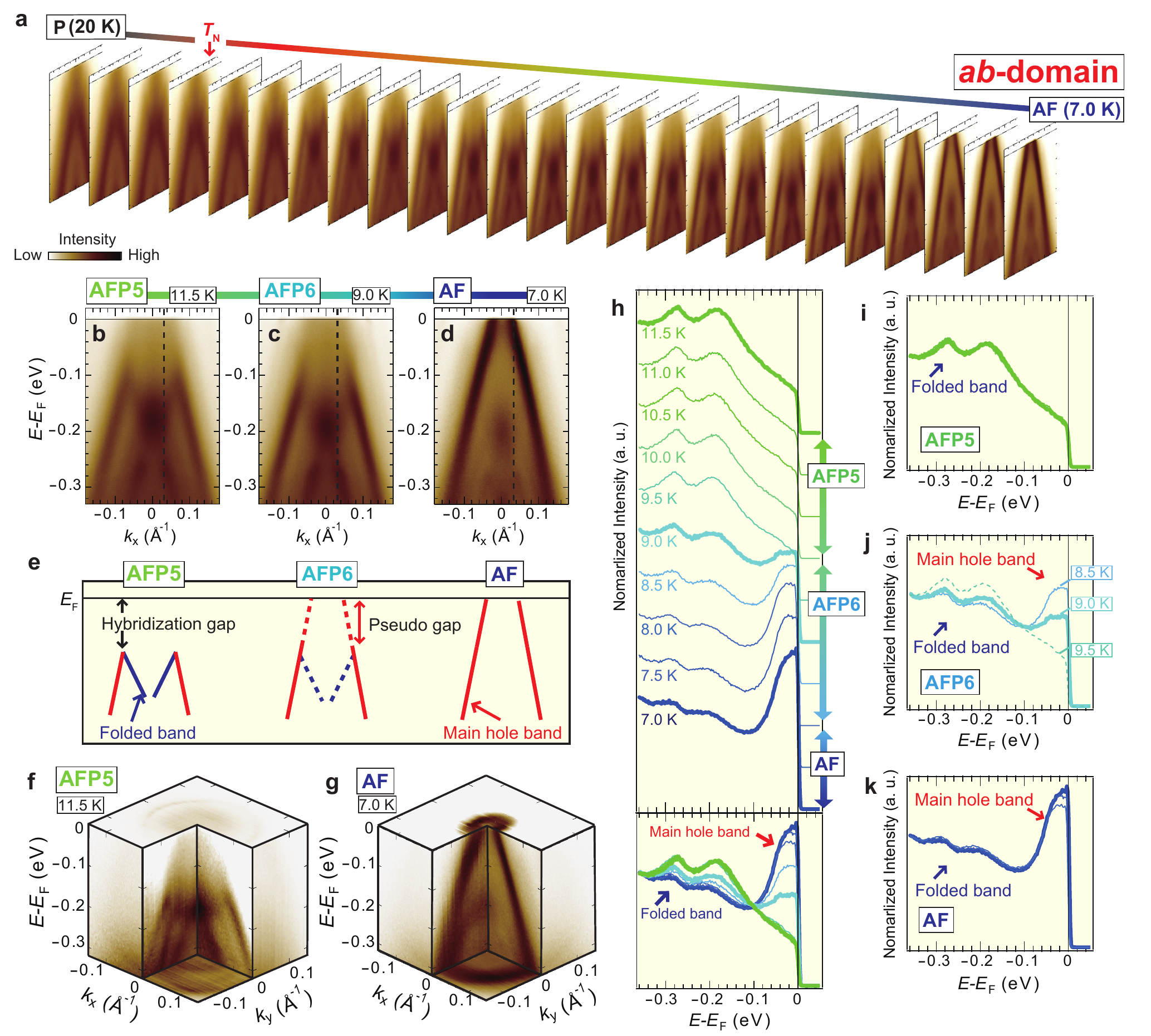}
\caption{\label{fig5}
\textbf{Collapses of the quasiparticle coherence at $E_{\rm{F}}$.}
(\textbf{a}) Temperature evolution of the laser-ARPES images for $ab$-domain from 20~K (P phase) to 7.0~K (AF phase) by a 0.5~K step (see also Supplementary movie~1).
(\textbf{b-d}) The laser-ARPES images around $E_{\rm{F}}$ at representative phases.
(\textbf{e}) The schematics of the observed evolution of (red line) the main hole band dispersion and (blue line) the folded band appeared around $E-E_{\rm{F}}$=0.12~eV in AFP5 phase.
(\textbf{f, g}) The three-dimensional ARPES images at 11.5 K (AFP5 phase) and 7.0~K for (AF phase), respectively. 
(\textbf{h}) Temperature evolution of the spectral shape in the energy distribution curves (EDCs) cut at a +$k_x$ point of the hole band at AF phase (black dashed line in \textbf{b-d}).
In the top and bottom panels, the data are displayed with and without an offset, respectively.
The representative EDCs corresponding to the different phases are highlighted in bold lines.
(\textbf{i-k}) The EDCs at various temperatures for each phase.
}
\end{figure*}

The significant reordering of the electronic structure together with the devil's staircase and upon the anisotropic 4$f$ modulation-field has been so far evaded from the experimental detection for a long-time despite the 40-years of intense researches.
Our high-quality data capture it and therefore provide the first opportunity of tracing the temperature evolution of electronic structure going through the devil's staircase.
We verify changes of the spectral response at each transition for $c$- and $ab$-domains by comparing the ARPES images to the long-periodicity developing through a number of the AFP transitions.
Eventually, the clear signature of the band folding obtained for $c$-domain substantiates the devil's staircase transition of the electronic band structures (see Fig.~\ref{fig4}).
We also reveal another important aspect of the devil's staircase state through the spectral observation of $ab$-domain: the band picture collapses by losing long-lived quasiparticle near $E_{\rm{F}}$ under the emergence of the long-periodic modulation in AFP6 phase (see Fig.~\ref{fig5}). 
We will present all of these observations below.\\

\noindent
{\textbf{The devil's staircase transition of the electronic bands.}}\\
Figure~\ref{fig4}\textbf{a} summarizes the temperature evolution of the laser-ARPES images for $c$-domain across $T_{\rm{N}}$, acquired with changing temperature with a 0.5~K step.
In Figs.~\ref{fig4}\textbf{b-i}, we present the representative images for all the different phases and the corresponding momentum distribution curves (MDCs).
At $T$=16.5~K (AFP1 phase, Fig.~\ref{fig4}\textbf{c}), the data displays the folded electron and hole bands (the dashed lines) in addition to the main hole band (the solid lines).
These bands begin to hybridize through $p$-$d$ mixing only with a slight decrease in temperature by 0.5~K (AFP2 phase at $T$=16.0~K, Fig.~\ref{fig4}\textbf{d}), transforming into the $M$-shaped dispersion.
As the 4$f$ modulation gets a longer periodicity with going from AFP2 to AFP6 phase, the hybridized band further splits in energy (colored lines in Figs.~\ref{fig4}\textbf{d}-\textbf{f}), and a number of the new bands appear.
Other transformations of the bands found in our laser-ARPES images are presented in Supplementary Fig.~4.
The dramatic evolution of the electronic structure is summarized also in Supplementary movie~1.  

In Figs.~\ref{fig4}\textbf{j-l}, the devil's staircase of the electronic structure is demonstrated as the temperature evolution of the MDCs at $E_{\rm{F}}$.
One can clearly see a series of quasiparticle peaks which dynamically appear and disappear as a function of temperature (Fig.~\ref{fig4}\textbf{j}).
This behavior can also be traced in the two-dimensional intensity map of the MDCs and its curvature plot~\cite{Zhang_rsi11} (Figs.~\ref{fig4}\textbf{k} and \ref{fig4}\textbf{l}, respectively).
For instance, the main peak for AFP3 phase is observed at $k_{x}\sim$0.14~$\rm{{\AA}}^{-1}$ only in a limited temperature range (orange arrow), and disappears at the transition to AFP4 phase.
Instead, the new peak appears at $k_{x}\sim$0.12~$\rm{{\AA}}^{-1}$ (green arrow) and at $\sim$0.11~$\rm{{\AA}}^{-1}$ for AFP5 phase (blue arrow).
The abrupt change of the spectra displays the temperature of each distinct transition in the devil's staircase, which almost matches with that determined by the previous specific heat measurement~\cite{Mignod_jpc1980} (horizontal solid lines in Figs.~\ref{fig4}\textbf{k} and \ref{fig4}\textbf{l}).

A major transition of the 4$f$ order occurs at $T_{\rm{AF}}\sim$8~K, in which the P layers of the AFP magnetostructures disappear in the whole crystal (see Fig.~\ref{fig1}\textbf{a}).
This impact can be observed in our data as the disappearance of quasiparticle peaks at $k_x\sim$0.14 and 0.17~$\rm{{\AA}}^{-1}$ (black circles in Figs.~\ref{fig4}\textbf{j} and~\ref{fig4}\textbf{l}), which corresponds to a contraction of the periodicity in the 4$f$ modulation from AFP6 to AF phase.
These results demonstrate that the band structure of the conducting electrons sensitively relies on the periodicity of the 4$f$ modulation.\\

\noindent
{\textbf{Spectral anomaly by the long-periodic modulation.}}\\
In contrast to the well-defined bands observed in $c$-domain (Fig.~\ref{fig4}), our high-quality data obtained in $ab$-domain reveal that the band picture collapses around $E_{\rm{F}}$ by losing quasiparticle peaks in AFP6 with the long-periodic modulation (Fig.~\ref{fig5}).
This indicates that not only the spectral peak positions (or energy states) but also their spectral weight are crucial quantities to investigate in fully understanding the electronic properties of the complex devil's staircase state.

Figure~\ref{fig5}\textbf{a} exhibits the temperature evolution of the laser-ARPES images for $ab$-domain across $T_{\rm{N}}$, acquired with changing temperature with a 0.5~K step (see also Supplementary movie~1).
The significant variation is found through the transitions from AFP5 to AF phase: the well-defined band disperses across $E_{\rm{F}}$, forming the main hole band in AF phase (Fig.~\ref{fig5}\textbf{d}), whereas the clear Fermi pocket is eliminated in AFP5 phase by the energy gap (hybridization gap) opened around $E_{\rm{F}}$ due to the hybridization between the main and folded bands, and instead the $M$-shaped band is formed below $E-E_{\rm{F}}=-$0.12~eV (Fig.~\ref{fig5}\textbf{b}).
This temperature evolution of the band structures is schematically illustrated in Fig.~\ref{fig5}\textbf{e}.
In Figs.~\ref{fig5}\textbf{f} and~\ref{fig5}\textbf{g}, we present band structures in three-dimensions for AFP5 and AF phase, respectively, by plotting the spectral intensity both for the FS surface and energy dispersion; these demonstrate that the opening of the hybridization gap widely takes place in the observed $k$ region under the magnetic modulation of a long-periodicity in AFP5 phase. 

To better understand the temperature evolution going from AFP5 to AF phase, we extract the energy distribution curves at a $k_{\rm{F}}$ point for AF phase in Fig.~\ref{fig5}\textbf{h}.
In the top and bottom panels, the spectra are plotted with and without an offset, respectively.
The spectral shape is drastically changed within a very narrow range of temperatures between 11.5~K and 7.0~K.
Significantly, the spectral weight of the quasiparticle peak near $E_{\rm{F}}$ at 7.0~K (blue lines, AF phase) is completely transferred to higher binding energies at 11.5~K (green lines, AFP5 phase), yielding a peak for another energy state of the folded band (a blue arrow).

We find that the significant spectral transfer occurs only in AFP6 phase whereas the spectral shape is almost unchanged with temperature in the other two phases (AFP5 and AF phase), as demonstrated from Fig.~\ref{fig5}\textbf{i} to \ref{fig5}\textbf{k} by separately overlapping spectra for each phase.
Consequently, an abnormal spectral feature is obtained in AFP6 phase: the spectral weight near $E_{\rm{F}}$ and that for the folded band becomes comparable as schematically illustrated in Fig.~\ref{fig5}\textbf{e} with dashed lines.
This unusual state in AFP6 phase manifests the pseudogap-like anomaly without the long-lived quasiparticle (indicated by a red arrow in the middle cartoon of Fig.~\ref{fig5}\textbf{e}), which substantially differs from the band reconstructions associated both with opening the hybridization gap in AFP5 phase (left cartoon of Fig.~\ref{fig5}\textbf{e}) and with forming a clear hole pocket in AF phase (right cartoon of Fig.~\ref{fig5}\textbf{e}).
Let us, however, remind that the well-defined bands are observable even in the AFP phases for $c$-domain, in which we cut the different momentum plane (Fig.~\ref{fig4}).
We thus conclude that the pseudogap-like anomaly without the well-defined band appears in a portion of the tetragonal FSs.

Our data clearly reveal that the emergence of the ordered 4$f$ states reorganizes not only the electronic structures but also spectral property, sensitively depending on the periodicity and symmetry of the 4$f$ modulation.
These experimental observations give a new electronic insight to the devil’s staircase beyond the general studies of the localized 4$f$ electrons, and strongly suggest that the devil's staircase nature results from the electronically driven instability.
In the spatially modulated systems, electronic instability is often observed as pseudo-gap behaviors in the ground state~\cite{Eefetov_naturephys2013, Kim_science2014, Comin_science2014}.
Unlike the typical cases, however, our data of ordered CeSb exhibit such a spectral anomaly in the transitional phase at temperatures ($T_{\rm{N}}>T>T_{\rm{AF}}$) slightly higher than that of the lowest-temperature phase ($T<T_{\rm{AF}}$), where the quasiparticle coherence is recovered below $T_{\rm{N}}$.
This unique property suggests the presence of electronic instabilities competing with each other in the transitional AFP phases, which is of importance for understanding the mechanism of the devil's staircase.
With these our results, we hope to simulate the advanced theories including the electronic correlation effect and the coupling with the ordered 4$f$ state, which is capable of explaining the observed spectral response from first principles.

Finally, we emphasize that the direct observation of the dramatic reconstruction previously refused by surface-sensitive ARPES~\cite{Kumigashira_prb1997,Ito_physica2004,Jangeaat_scienceAd2019} has now become possible owing to the advantage of low $h\nu$ source, which allows the acquisition of high-quality data to precisely determine the bulk electronic state.
Since the magnetostructure phase transition of CeSb presents the most complex phenomena among those of Ce monopnictides CeX (X: P, As, Sb or Bi)~\cite{Kohgi_physica2000}, it is rather interesting to systematically investigate all the CeX by ARPES with a low $h\nu$ laser, and compare the low-energy electronic structures in the ordered phases of these compounds.
Such systematic investigation would give a great insight into the electronically competing states causing the devil's staircase.

\noindent\\
{\textbf{Methods}\\}
{\textbf{Sample growth.}\\}
CeSb single crystals were grown by Brigeman method with a sealed tungsten crucible and high-frequency induction furnace.
High-purity Ce (5N) and Sb (5N) metals with the respective composition ratio were used as starting materials.
The obtained samples were characterized by the DebyeScherrer method.
\\
\noindent\\
{\textbf{Polarized microscopy measurements.}\\}
The polarizing image measurement was performed at ISSP, the University of Tokyo~\cite{Katakura_rsi2010}.
The sample temperature was controlled in the range of 8-20~K.
A 100~W halogen lamp (U-LH100L-3, Olympus) was used to obtain bright images.
The flat (001) surfaces for imaging were prepared by cleavage in atmosphere, and the sample was immediately installed to the vacuum chamber within 10 minutes.
The polarizing microscope images were taken in crossed Nicols configuration with the optical principal axes along [110].
\\
\noindent\\
{\textbf{Angle-resolved photoemission experiments.}\\}
The high-resolution laser-ARPES was performed at ISSP, the University of Tokyo~\cite{ShimojimaJPSJ2015}.
The energy (angular) resolution was set to 2~meV (0.3$^{\circ}$).
The temperature stability during each scan was better than $\pm$0.1~K.
A Scienta-Omicron R4000 hemispherical electron analyzer was used, with a vertical entrance slit and the light incident in the horizontal plane.
The linear polarization with the electric field aligned on the incident plane was used. 
The synchrotron radiation ARPES experiments with tunable low-$h\nu$ were performed at the BL-9A beamline of Hiroshima Synchrotron Radiation Center (HiSOR) and the BL7U beamline of UVSOR-III.
The experiments were carried out with Scienta-Omicron R4000 analyzer (HiSOR) and MBS A-1 analyzer (UVSOR), respectively.
The energy-resolution was better than 15~meV and the angular-resolution was 0.3$^{\circ}$.
In all of these ARPES experiments, the base pressure in the chamber was better than 1$\times$10$^{-8}$~Pa.
The crystals were cleaved $in\; situ$ at the measurement temperature ($T$=10~K).
For the measurements of the temperature dependence, the ARPES images were scanned from the lower to higher temperatures.  
\\
\noindent\\
{\textbf{Calculations.}\\}
The DFT band structure calculations were performed using the VASP package~\cite{Kresse_prb96,Kresse_cms96}, with the experimental lattice constant.
All of calculations were done within non-spin polarized approximation with spin-orbit correction (SOC).
To approximate electron exchange-correlation energy, GGA functional was used.
In addition, the Hubbard correction $U$ was employed within DFT+U method to improve the electron-electron repulsion for Ce 5$d$ electrons.
The large $U$ value (7~eV) was necessary used to reproduce the bulk band dispersions, particularly the energy position of the 5$d$ band bottom, observed by our previous soft x-ray ARPES~\cite{Kuroda_prl2018}.
The Kohn-Sham equation was solved through PAW method and the wave functions were expanded by a plane wave basis set with a cutoff energy of 700~eV.
The Ce 4$f$-electrons were treated as core electrons in all of the calculations.
The integration over the Brillouin zone was done using 16$\times$16$\times$16 Monkhorst-pack mesh.

We also have conducted spin-polarized calculation including 4$f$ electrons into valance states, and the result is compared to that with non-spin polarized calculations (see Supplementary Fig.~5). 
In this part of the calculations, we have to apply Hubbard correction into 4$f$ electrons within the DFT+$U$ method otherwise, 4$f$ related states wrongly will appear at $E_{\rm{F}}$.
For both of these calculations ($i.e.$ spin-polarized and non-spin-polarized), we use full-potential linearized augmented plane-wave (FLAPW) implemented in Fleur code.
Because in DFT+$U$, we are allowed to apply Hubbard U into only one type of orbital per atom, and since in spin-polarized calculation we have to use Hubbard correction for 4$f$ electrons, therefore to make a sensible comparison between spin-polarized and non-spin polarized.
By these reasons, we cannot apply $U$ into 5$d$ electron of Ce in both cases and thus the energy position of the Ce 5$d$ band differs from the experimental one.
\\
\noindent\\
{\textbf{Data Availability}\\}
The data that support the findings of this study are available from the corresponding authors upon reasonable request.
\\
\\
\noindent\\

%
\noindent\\
{\textbf{Acknowledgements}}\\
We acknowledge O.~Sakai, K.~Iwasa, T.~Osakabe, Y.~Tada, T.~Shimozawa and H.~Kusunose for fruitful discussions and T.~Hashimoto, T.~Tsuzuki, Y.~Ota, Y.~Ishida for supports of experiments.\\
\noindent\\
\textbf{Author Contribution}\\
K.Ku. and Y.A. conducted laser ARPES experiments and analyzed the data.
S.K., S.S., C.B., R.N., M.N., S.A., M.S., K.Ka., M.A., S.I., K.T., K.O., S.S., T.K. supported ARPES experiments.
H.S.S., H.K. made high-quality CeSb single crystals.
K.Ku., Y.A., Y.K., M.T. performed polarizing microscopy.
N.R., M.A., R.A. calculated the band structures.
H.S.S., Y.H., R.A. provided theoretical insights.
K.Ku., T.K. wrote the paper.
All authors discussed the results and commented on the manuscript.\\
\noindent\\
\textbf{Correspondence}\\
Correspondence and requests for materials
	should be addressed to K.Kuroda~(email: kuroken224@issp.u-tokyo.ac.jp).

\clearpage
%
\clearpage
%
%
\clearpage
%
\clearpage
%
%

\end{document}